\documentclass[
prc,%
10pt,%
final,%
notitlepage,%
oneside,%
twocolumn,%
nobibnotes,%
nofootinbib,
superscriptaddress,%
floatfix,%
floatfix,%
showkeys,%
showpacs]%
{revtex4}
\usepackage{color}
\usepackage{amsfonts}
\usepackage{amsbsy}
\usepackage{mathrsfs}
\usepackage{graphicx}
\def\lsim{\mathrel{\rlap{
\lower4pt\hbox{\hskip-3pt$\sim$}}
    \raise1pt\hbox{$<$}}}     
\def\gsim{\mathrel{\rlap{
\lower4pt\hbox{\hskip-3pt$\sim$}}
    \raise1pt\hbox{$>$}}}     
\def\scr#1{\mbox{\scriptsize #1}}
\begin{document}
\title{
Light fragment production at CERN Super Proton Synchrotron} 
\author{Yu. B. Ivanov}\thanks{e-mail: y.b.ivanov@yandex.ru}
\affiliation{National Research Centre "Kurchatov Institute", 123182 Moscow, Russia} 
\affiliation{National Research Nuclear University MEPhI  (Moscow Engineering
Physics Institute),
Moscow 115409, Russia}
\affiliation{Bogoliubov Laboratory of Theoretical Physics, JINR, 141980 Dubna, Russia}
\author{A. A. Soldatov}\thanks{e-mail: saa@ru.net}
\affiliation{National Research Nuclear University MEPhI  (Moscow Engineering
Physics Institute),
Moscow 115409, Russia}
\begin{abstract}
Recent data   
on the deutron and $^3$He production in central Pb+Pb collisions at the CERN Super Proton Synchrotron (SPS)
energies measured by the NA49 Collaboration are analyzed  within the model of the three-fluid dynamics
(3FD) complemented by the coalescence model for the light-fragment production. 
The simulations are performed with different 
equations of state---with and without deconfinement transition.
It is found that scenarios with the deconfinement transition are preferable 
for reproduction rapidity distributions  of deuterons and $^3$He, the corresponding results well agree 
with the experimental data.
At the same time the calculated 
transverse-mass spectra 
at midrapidity do not that nice agree 
with the experimental data. 
The latter apparently indicates that coalescence coefficients should be temperature and/or 
momentum dependent.
\end{abstract}
\maketitle

\section{Introduction}

Recently experimental data on light-fragment production in Pb+Pb collisions at SPS energies 
has been published by the NA49 Collaboration \cite{Anticic:2016ckv}.  
These data has been already theoretically analyzed in Refs.  
\cite{Anticic:2016ckv,Sun:2017xrx,Mrowczynski:2016xqm}. 
Traditionally the light-fragment data are interpreted within either the thermodynamical or coalescence 
models which in fact give results quantitatively close to each other \cite{Mrowczynski:2016xqm}.
The above-mentioned approaches are based on schematic fireball-like models which analyze total 
(or midrapidity) yields of light fragments. 
Nevertheless, the NA49 data include spectra in a wide range of rapidity and transverse 
momentum rather then only total yields. 
In the present study we would like to focus on the coalescence approach and 
to address the questions: 
\\
(i) If the coalescence within 3D simulations is able to reproduce the 
rapidity and transverse-momentum spectra 
of light fragments rather than only their total (or midrapidity) multiplicities? 
\\
(ii) If these spectra are sensitive to the equation of state (EoS) used in the simulations, in particular, 
to the deconfinement transition?

In the present paper the Pb+Pb collisions are simulated within the 3FD model  
\cite{3FD} 
for several collision energies in the SPS energy range. 
The 3FD model is quite successful in reproduction 
of the major part of bulk
observables in this range, among those the proton rapidity \cite{Ivanov:2013wha}  
and transverse-momentum distributions \cite{Ivanov:2013yla} 
are relevant to the present study. 
Light fragment formation (deutrons, tritons, $^3$He and $^4$He) is
taken into account in terms of the coalescence model, which is similar
to that described in  Appendix E of Ref. \cite{gsi94}.

\section{Coalescence in the 3FD model}
\label{Model}

Unlike the conventional hydrodynamics, where local
instantaneous stopping of projectile and target matter is
assumed, a specific feature of the 3FD description \cite{3FD} 
is a finite stopping power resulting in a counterstreaming
regime of leading baryon-rich matter. This generally
nonequilibrium regime of the baryon-rich matter
is modeled by two interpenetrating baryon-rich fluids 
initially associated with constituent nucleons of the projectile
 and target nuclei. In addition, newly produced particles,
populating the midrapidity region, are associated with a fireball
 fluid.
Each of these fluids is governed by conventional hydrodynamic equations 
coupled by friction terms in the right-hand sides of the Euler equations. 
These friction terms describe energy--momentum loss of the 
baryon-rich fluids. 
A part of this
loss is transformed into thermal excitation of these fluids, while another part 
gives rise to particle production into the fireball fluid.

The physical input of the present 3FD calculations is described in
Ref.~\cite{Ivanov:2013wha}. The friction between fluids was fitted to reproduce
the stopping power observed in proton rapidity distributions for each EoS, 
as it is described in  Ref. \cite{Ivanov:2013wha} in detail.
The simulations in \cite{Ivanov:2013wha,Ivanov:2013yla} 
were performed with different 
EoS's---a purely hadronic EoS \cite{gasEOS}  
and two versions of the EoS involving the   deconfinement
 transition \cite{Toneev06}, i.e. a first-order phase transition  
and a smooth crossover one. In the present study we use precisely  
the same parameters as those reported in Ref.~\cite{Ivanov:2013wha}.

The 3FD model does not describe clustering in the baryonic matter. 
Therefore, we need extra assumptions to calculate the fragment production. 
We describe the fragment production within a coalescence model,
in the spirit of refs.
\cite{Gutb76,Goss77,Kap80}, similar to that it was done in Ref. \cite{gsi94}. 
We assume that $N$ neutrons and $Z$ protons, falling within a
6-dimensional phase volume
$(\frac{4}{3} \pi p_{NZ}^3) (\frac{4}{3} \pi r_{NZ}^3)$
at the freeze-out stage, form a $(N,Z)$-fragment. Here
$p_{NZ}$ and $r_{NZ}$ are the parameters of the coalescence
model, which are, in principle, different for different
$(N,Z)$-fragments.
Sometimes the coalescence is performed only in the momentum space, 
when information on the configuration space is unavailable. 
Such kind of the coalescence was reported, e.g., in the paper 
of the NA49 Collaboration \cite{Anticic:2016ckv} and the above-mentioned
analysis of these data \cite{Sun:2017xrx,Mrowczynski:2016xqm}. 
However, from the physical point of view the nucleons should be close 
in both momentum and configuration spaces in order to be able to 
merge into a fragment. Therefore, in this paper we apply the local 
version of the coalescence, i.e. within the 6-dimensional phase space.

The consideration below concerns a single cell in the configuration space. 
To avoid multiple subscripts in the notation we suppress the cell subscript. 
We calculate the distribution of
observable $(N,Z)$-fragments as follows
(cf. \cite{Goss77,Gupt81})  
%
\begin{eqnarray}
\label{dN(NZ)/dp(a)}
&&E_A \; \frac{d^3 \tilde{N}_{N,Z}}{d^3 P_A} =
\frac{N_{tot}^N Z_{tot}^Z}{A_{tot}^A} \;
A \;
\frac{(\frac{4}{3} \pi p_{NZ}^3 / M_N)^{A-1}}{N! Z!} \;
\cr
&&\times 
\left(
\frac{V_{NZ}}{V}
\right)^{A-1} \;
\left(
E \; \frac{d^3 \tilde{N}^{(N)}}{d^3 p}
\right)^{A}, 
\end{eqnarray}
%
where 
$d^3 \tilde{N}^{(N)}/d^3 p$ is the distribution of
observable nucleons. 
Here
$N_{tot}=N_p+N_t$,
$Z_{tot}=Z_p+Z_t$ and
$A_{tot}=A_p+A_t$
are the total numbers of neutrons, protons and nucleons in the
projectile-plus-target nuclei, respectively, $A=N+Z$,
$E_A=AE$,
${\bf P}_A=A{\bf p}$, 
$V_{NZ}=\frac{4}{3} \pi r_{NZ}^3$,
and $M_N$ is the nucleon mass. 
$V=\bar{A}_{cell}/n_c$ is the total volume of the frozen-out
cell,
where $n_c$ is the freeze-out density and
%
\begin{equation}
\label{A(tot)}
\bar{A}_{cell} =
\int d^3 p
\; \frac{d^3 N^{(N)}}{d^3 p}
\end{equation}
%
is the total number of primordial participant nucleons.
Here we denote the distributions of observable (i.e. after the
coalescence) nucleons and fragments by a tilde
sign, in contrast to
the primordial nucleon distribution.
Unlike refs. \cite{Goss77,Gupt81}, we formulate the
coalescence in terms of invariant distributions
$E \; d^3 N / d^3 p$ and also introduce the factor
$(V_{NZ}/V)$, taking into account a vicinity in the coordinate
space. This factor 
makes $E_A \; d^3 \tilde{N}_{N,Z}/d^3 P_A  \sim V$, 
i.e. an additive quantity suitable for the summation over cells. 
Defining a new parameter
%
\begin{equation}
\label{P(NZ)}
P_{NZ}^3 =
\frac{4}{3} \pi p_{NZ}^3 \;
V_{NZ} \; n_c \;
\left(
\frac{A}{N! Z!} \;
\right)^{1/(A-1)},
\end{equation}
%
we can write down eq. (\ref{dN(NZ)/dp(a)}) in a simpler form
%
\begin{equation}
\label{dN(NZ)/dp(b)}
E_A \; \frac{d^3 \tilde{N}_{N,Z}}{d^3 P_A} =
\frac{N_{tot}^N Z_{tot}^Z}{A_{tot}^A} \;
\left(
\frac{P_{NZ}^3}{M_N \bar{A}_{cell}}
\right)^{A-1}
\left(
E \; \frac{d^3 \tilde{N}^{(N)}}{d^3 p}
\right)^{A},
\end{equation}
%
where 
$d^3 {N}^{(N)}/d^3 p$ is the distribution of
observable nucleons, i.e. those after the coalescence.  
In this form the fragment distribution contains only a single phenomenological 
parameter,  $P_{NZ}$, that defines the total normalization of the distribution. 
These equations for different $N$ and $Z$ form a set of equations,
since the nucleon distribution in the r.h.s. is an observable
distribution rather than a primordial one. To make this system
closed, one should add a condition of the baryon  number conservation
%
\begin{equation}
\label{N-conserv.}
E \; \frac{d^3         N^{(N)}}{d^3 p} =
E \; \frac{d^3 \tilde{N}^{(N)}}{d^3 p} +
\sum_{N,Z \; (A>1)} \; A^3 \;
E_A \; \frac{d^3 \tilde{N}_{N,Z}}{d^3 P_A}.
\end{equation}
%
Thus calculated distribution of observable fragments is summed over all cells 
in order to obtain the total momentum distribution of fragments.  
The $P_{NZ}$ parameters are fitted to reproduce normalization of 
spectra of light fragments. 

\section{Results}
\label{Results}

\begin{figure*}[!bt]
\includegraphics[width=15.5cm]{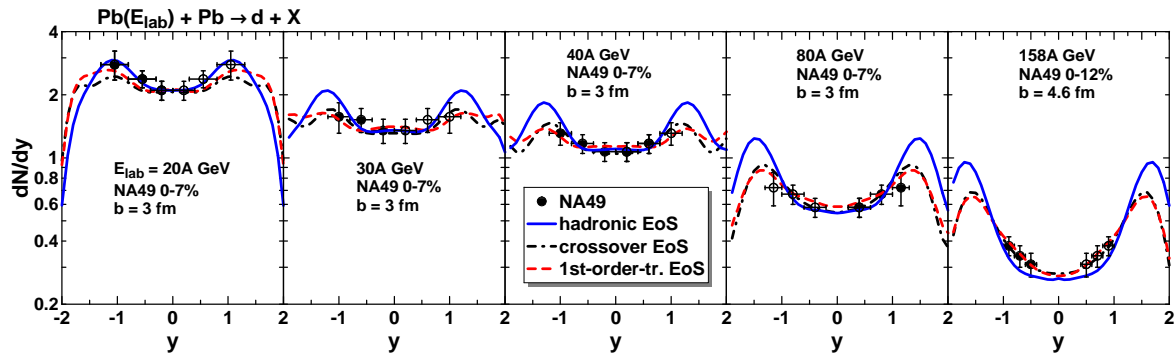}
 \caption{
Rapidity distributions of deuterons in central Pb+Pb collisions at various SPS energies ($E_{\rm lab}$) 
confronted to 3FD calculations with different EoS's. Experimental data are from the NA49 Collaboration
\cite{Anticic:2016ckv}. The percentage indicates the fraction of the total
reaction cross section, corresponding to experimental selection of events.
The solid
symbols show the measurements and the open symbols represent the data points reflected about mid-rapidity.
}
\label{fig1}
\end{figure*}
\begin{figure*}[!bt]
\includegraphics[width=15.5cm]{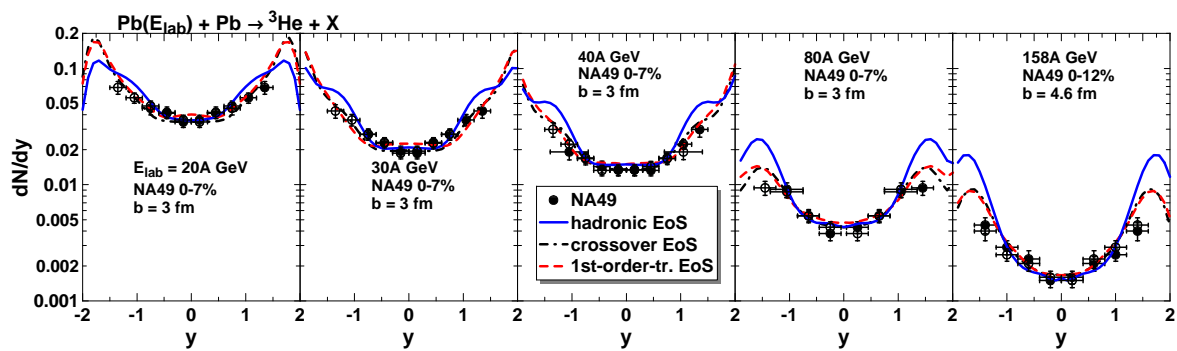}
 \caption{
The same as in Fig. \ref{fig1} but for $^3$He.
}
\label{fig2}
\end{figure*}

Table \ref{tab:2} presents results of the fit of the $P_{NZ}$ parameters to the 
NA49 data \cite{Anticic:2016ckv}. There is a clear trend of $P_{NZ}$ reduction with 
collision energy rise. This can be associated with properties of the freeze-out 
procedure adopted in the 3FD model \cite{Russkikh:2006aa,Ivanov:2008zi}. The freeze-out locally starts 
when the local energy density drops below some value (0.4 GeV/fm$^3$ in the present 
simulations). The thermal part of the energy density increases  with the 
collision energy rise. Therefore, the compressional part drops, so does the freeze-out 
baryon density [$n_c$, see Eq. (\ref{P(NZ)})]. This trend is less visible in the 
case of $^3$He. 

\begin{table}[htb]
\begin{ruledtabular}
  \begin{tabular}{|cc|ccccc|}
$E_{\scr{lab}}$ &[$A\cdot$GeV]         & 20 & 30 & 40 & 80 & 158  \\
$P(\rm{d})$ &[MeV/c]                   & 513& 471& 466& 431& 425  \\
$P(^3\rm{He})$ &[MeV/c]                & 474& 453& 449& 415& 409  \\
$\langle n_c/n_0\rangle $ &            & 0.61& 0.61& 0.57& 0.48& 0.43 \\
$P({\rm d})\langle n_c/n_0\rangle^{-1/3}$&[MeV/c]    & 606& 563& 562& 550& 563 \\
$P(^3{\rm He})\langle n_c/n_0\rangle^{-1/3}$&[MeV/c] & 559& 534& 542& 530& 542 \\
  \end{tabular}
\caption{Coalescence parameters, see Eq. (\ref{P(NZ)}),    
used in 3FD simulations of Pb+Pb
collisions at various incident energies $E_{\scr{lab}}$, 
the corresponding mean baryon densities ($n_c$) at the freeze-out divided
by normal nuclear density ($n_0$) calculated within the crossover scenario,  
and the reduced parameters $P_{NZ}\langle n_c/n_0\rangle^{-1/3}$. 
}
\label{tab:2}
\end{ruledtabular}
\end{table}

The above mentioned decrease of the freeze-out baryon density is illustrated in Table \ref{tab:2}. 
The displayed mean freeze-out baryon density  is calculated within the crossover scenario
for Pb+Pb collisions at impact parameters $b=$ 2.4 fm for $E_{\scr{lab}}=$ 20$A$-80$A$ GeV and 
$b=$ 4.6 fm for $E_{\scr{lab}}=$ 158$A$ GeV, which correspond the experimental centrality selection. 
In order to remove the $n_c$ dependence from the $P_{NZ}$ parameters, these parameters were reduced 
to the normal nuclear density ($n_0=$ 0.15 fm$^{-3}$): $P_{NZ}\langle n_c/n_0\rangle^{-1/3}$, cf. Eq. (\ref{P(NZ)}). 
As it is seen from Table \ref{tab:2}, these reduced parameters are constant in the considered energy range 
with the accuracy of $\sim$2\%, except for the case of the collision energy of 20$A$ GeV. 
Enhanced production of light fragments can be a signature of onset of the first-order 
phase transition at 20$A$ GeV: the density fluctuations in the spinodal region can manifest 
themselves this way. 
The data on tritons were not analyzed because their experimental accuracy is much lower than 
that for deuterons and $^3$He.

The deduced $P_{NZ}$ parameters may look too high from the point of view of the 
coalescence performed only in the momentum space. Though, the $P_{NZ}$ parameters 
are effective momenta, cf. Eq. (\ref{P(NZ)}). Only $p_{NZ}$ momenta have a clear physical meaning. 
We are not able to separately determine the $p_{NZ}$  and $V_{NZ}$ parameters. 
However, we can estimate $V_{NZ}$. The number of nucleons  
in the $V_{NZ}$ volume  should not be less than two the for the case of the deuteron
(or three for $^3$He), i.e. 
$V_{NZ} \; n_c \gsim 2$ (or 3 for $^3$He). 
Then collecting all factors in Eq. (\ref{P(NZ)}), we see that the $p_{NZ}$ values are at least 
2.5 times or more lower then $P_{NZ}$ ones. These are already reasonable values for the 
coalescence which are well comparable with those used in the Quark-Gluon String Model (QGSM)
\cite{Toneev:1983cb,Steinheimer:2012tb}. In particular, if we accept the QGSM recipe \cite{Toneev:1983cb,Steinheimer:2012tb}, i.e. $p_{NZ}\; r_{NZ} = \hbar$, we arrive at 
$P^{QGSM}(d) = 512$ MeV/c for deuterons which is close to the values presented in Table \ref{tab:2}.

Results for the rapidity distributions of deuterons and $^3$He in central Pb+Pb collisions 
at various SPS energies are presented in Figs. \ref{fig1} and \ref{fig2}, respectively. 
As a rule, scenarios with deconfinement transition perfectly reproduce the NA49 data 
\cite{Anticic:2016ckv}. The hadronic scenario looks preferable only for deuterons at 
20$A$ GeV. We would like to remind that these results are achieved with only a single 
parameter for each distribution which determines the overall normalization of the spectrum. 
Correspondence between the fraction of the total reaction cross
section related to a data set and a mean value of the impact
parameter ($b$ in Figs. \ref{fig1} and \ref{fig2}) was read off from the paper \cite{Alt:2003ab}.
Of course, this correspondence is only approximate because in fact the experimentally selected 
events populate a certain range of the impact parameters rather than are related a single $b$.  
Moreover, the $b=4.6$ fm bin corresponds to centrality from 5\% to 12.5\%, while the bin of 
Ref. \cite{Anticic:2016ckv} for the energy 158$A$ GeV contains events with 0-12\% centrality. 
Nevertheless, we use the $b=4.6$ fm impact parameter as an approximate representative of the 
0-12\% bin.

\begin{figure}[tbh]
\includegraphics[width=7.cm]{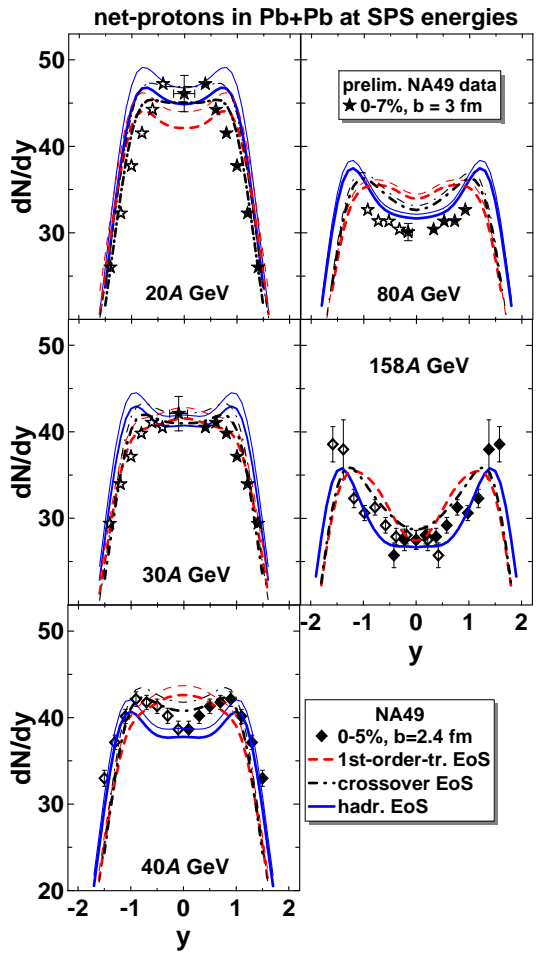}
 \caption{
Rapidity distributions of net-protons in central Pb+Pb collisions at SPS energies
calculated within three considered scenarios. 
Thin lines display results without subtracting the contribution of light 
fragments, these were earlier reported in Ref. \cite{Ivanov:2013wha}.
Bold lines present the results corrected by subtracting the contribution of light 
fragments.
Experimental data
are from the NA49 collaboration \cite{NA49-1,NA49-04,NA49-06,NA49-07,NA49-09}. 
The percentage shows the fraction of the total
reaction cross section, corresponding to experimental selection of events.
}
\label{fig3}
\end{figure}
\begin{figure}[tbh]
\includegraphics[width=5.cm]{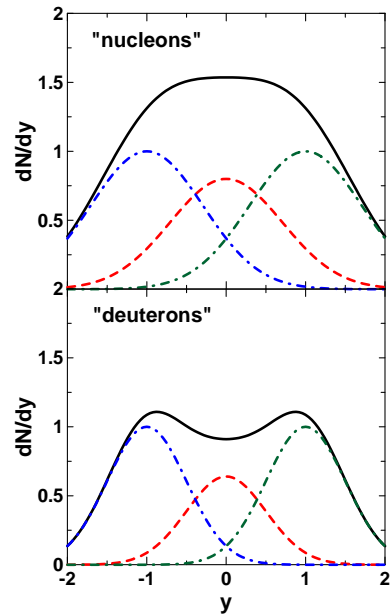}
 \caption{
Schematic illustration of rapidity distributions of protons (upper panel) and deuterons
(lower pannel), cf. Eqs. (\ref{N-toy}) and (\ref{d-toy}).  
}
\label{fig3.4}
\end{figure}
%
\begin{figure*}[!bht]
\includegraphics[width=15.5cm]{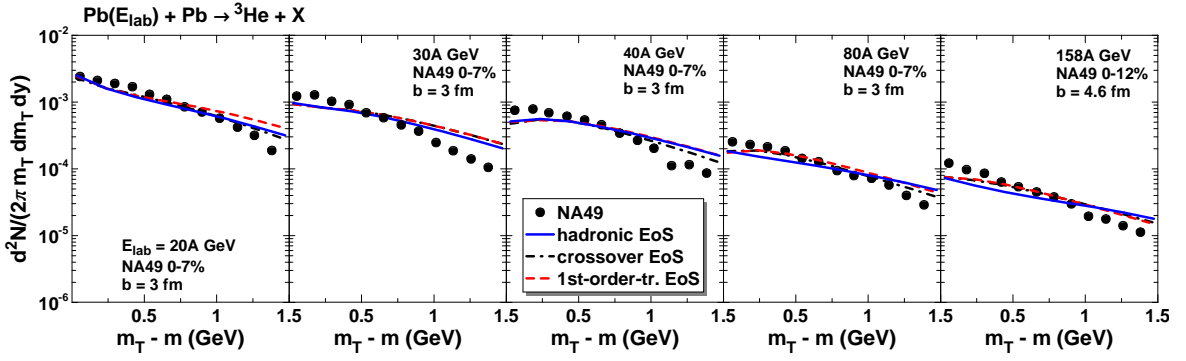}
 \caption{
Transverse-mass spectra of  $^3$He at midrapidity 
in central Pb+Pb collisions at various SPS energies ($E_{\rm lab}$) 
confronted to 3FD calculations with different EoS's. Experimental data are from the NA49 Collaboration
\cite{Anticic:2016ckv}. The percentage indicates the fraction of the total
reaction cross section, corresponding to experimental selection of events.
}
\label{fig4}
\end{figure*}
\begin{figure}[!tbh]
\includegraphics[width=6.cm]{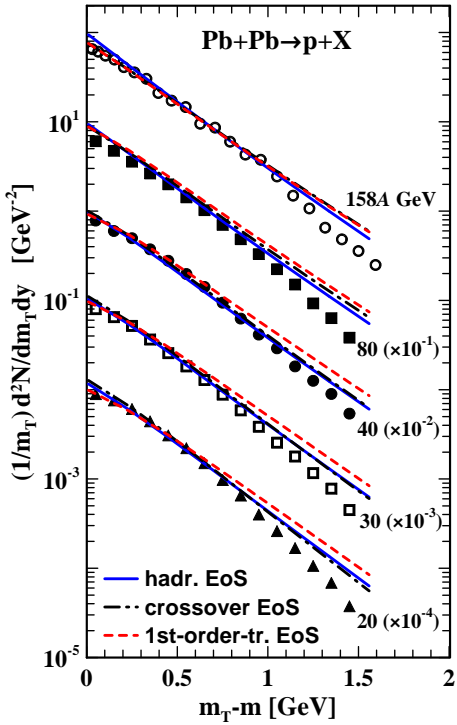}
 \caption{
Transverse mass spectra (at midrapidity) of protons from
central Pb+Pb collisions at SPS energies. Experimental data
are from the NA49 collaboration \cite{NA49-1,NA49-04,NA49-06,NA49-07,NA49-09}. 
}
\label{fig5}
\end{figure}
\begin{figure}[!tbh]
\includegraphics[width=6.cm]{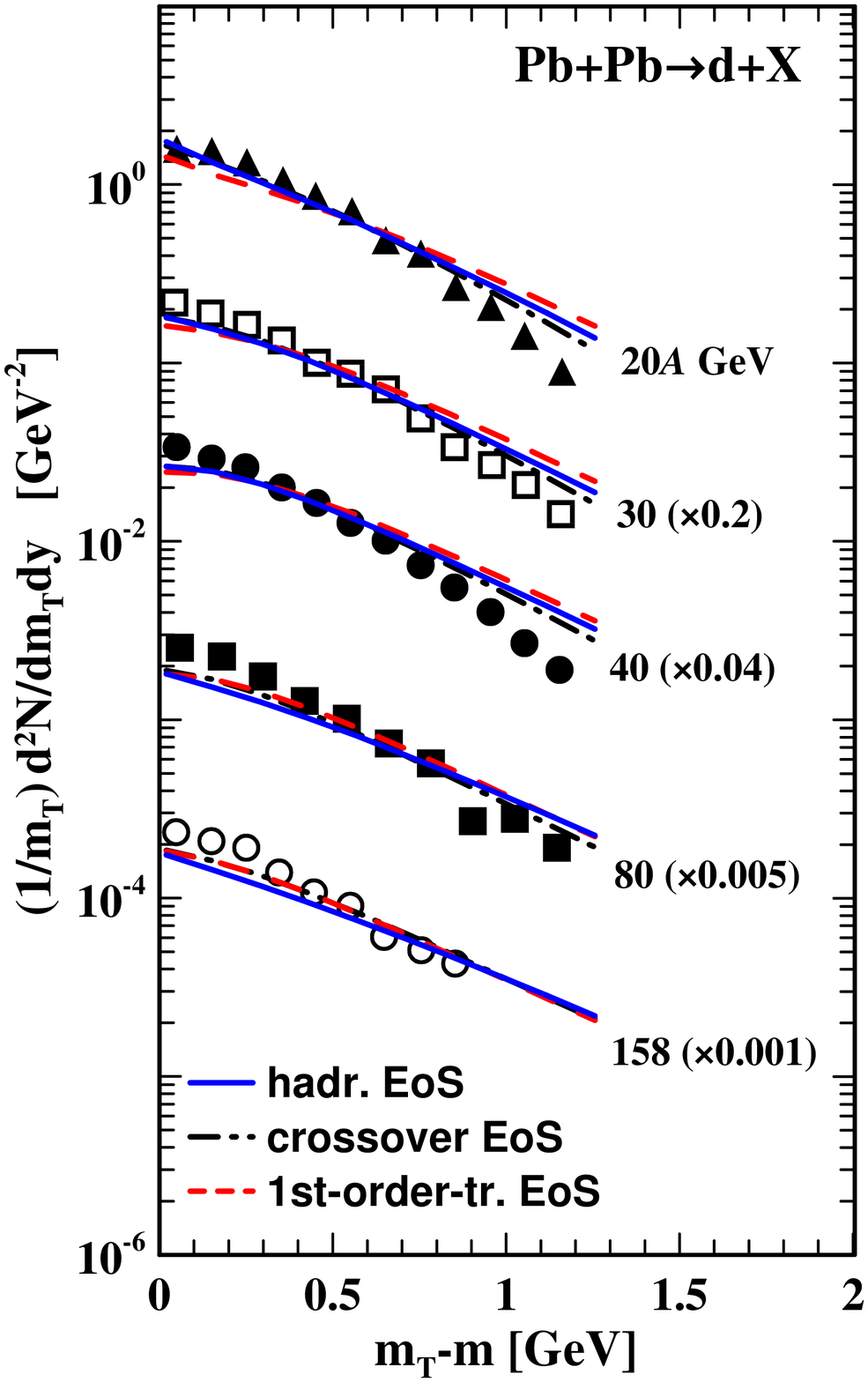}
 \caption{
The same as in Fig. \ref{fig4} but for deuterons. 
}
\label{fig6}
\end{figure}

Protons bound in light fragments should be subtracted from the calculated proton yield in 
order to compare the latter with observable proton data. 
At lower SPS energies this light-fragment correction is sizable. 
Rapidity distributions of net-protons in central Pb+Pb collisions at SPS energies
calculated with and without this correction are presented in Fig. \ref{fig3}. 
The net-proton distributions without the light-fragment correction were 
earlier reported in Ref. \cite{Ivanov:2013wha}. 
The light-fragment correction is indeed noticeable at 20$A$ GeV and improves 
agreement of the crossover results with experimental data. 
Note however that the data at 20$A$ GeV and  30$A$ GeV still have preliminary status, 
and hence it is too early to draw any solid conclusions 
from comparison with them. The light-fragment correction at 158$A$ GeV 
is practically negligible.

The dips in rapidity distributions of the fragments at the midrapidity are deeper than 
those in the corresponding proton distributions. In two cases, i.e. at 30 and 40$A$ GeV in the 
first-order-transition scenario, the midrapidity peak in the proton distributions transforms into 
the midrapidity dip in the fragment distributions. This is an effect of the local version 
of the coalescence applied here, i.e. the coalescence in the 6D phase space. Let us consider 
a very simplified toy example. Consider a three sources 
contributing to the midrapidity distributions (see Fig. \ref{fig3.4}): 
%
\begin{equation}
\label{N-toy}
\frac{dN_N}{dy} = \exp[-(y-1)^2]+ 0.8\exp[-y^2]+\exp[-(y+1)^2]. 
\end{equation}
%
These three sources are associated with three different spatial cells of the system 
rather than with the three different fluids 
inherent in the 3FD. Moreover, these fluids are mostly unified at the freeze-out stage. 
The number of cells in our calculations amounts to several hundreds of thousands.  
In this toy nucleon distribution we consider only three of these cells just to illustrate 
their interplay at the local coalescence. 
We disregard the transverse momentum as being irrelevant 
to our speculation. 

Thus, these  sources are composed of the same already unified fluid 
and are located in different places 
of the configuration space: the central source, $0.8\exp[-y^2]$, and two sources located 
closer to the target and projectile fragmentation regions, $\exp[-(y\pm 1)^2]$. 
It is important that the strength of the central source is smaller than those of the peripheral sources. 
In spite of this smallness, we have a midrapidity peak in the nucleon distribution 
because peripheral sources also contribute to the midrapidity nucleon yield due to 
a thermal spread in rapidity, simulated by the Gaussian form of the sources, see upper 
panel of Fig. \ref{fig3.4}.

After application of the local coalescence to the distribution (\ref{N-toy}), 
i.e. separately to each source, 
the deuteron rapidity distribution reads 
%
\begin{equation}
\label{d-toy}
\frac{dN_d}{dy} = (\exp[-(y-1)^2])^2+ (0.8\exp[-y^2])^2+(\exp[-(y+1)^2])^2. 
\end{equation}
%
The sum of these three sources already manifests a dip at the midrapidity 
because the strength of the central source is now relatively smaller than that 
in the nucleon distribution (\ref{N-toy}) and the 
thermal rapidity spread of the deuteron sources is lower than that in the nucleon distributions. 
Thus, the midrapidity peak in  the nucleon rapidity distribution transforms into 
the midrapidity dip in the deuteron rapidity distribution. 
Note that no such transformation happens if the coalescence is applied only in the 
momentum space, i.e. directly to the sum of the three sources in Eq. (\ref{N-toy}).

Figure \ref{fig4} presents the comparison of transverse-mass spectra of  $^3$He at midrapidity 
with experimental data from Ref. \cite{Anticic:2016ckv}. Here the agreement with the data 
is not that nice as that for the rapidity distributions. Moreover, different scenarios (with and 
without the deconfinement transition) fail approximately to the same extent. 
At the same time, within all considered scenarios 
the midrapidity transverse-mass spectra of protons are well reproduced 
in the low-$m_T$ range \cite{Ivanov:2013yla}, see Fig. \ref{fig5},  
which is relevant to the present light-fragment data. 

The reproduction of the data on the transverse-mass spectra
\cite{Anticic:2016ckv} is somewhat better in the case of 
deuterons, see  Fig. \ref{fig6}. However, the same trend of disagreement survives also 
for deuterons. Because of the worse statistics, the deuteron data correspond to the rapidity 
window $\Delta y \approx 0.6$ \cite{Anticic:2016ckv}. Therefore, for the comparison with 
these data we took calculations at $y = 0.3$. Though, the $y$-dependence of the 
deuteron transverse-mass spectra in this rapidity window is rather weak, as it results from 
the simulations. 

Poor reproduction of the transverse-momentum spectra indicates that the constant coalescence coefficient $(\frac{4}{3} \pi p_{NZ}^3) (\frac{4}{3} \pi r_{NZ}^3)$ is not the best choice. Of course, the coalescence coefficient can depend on the fragment momentum and also on the local temperature and baryon density. However, in the spirit of the local coalescence applied here the momentum should be taken in the local rest frame of the fluid element, i.e. $u_\mu p^\mu$ where $u_\mu$ is the local 4-velocity of the fluid. The dependence on the baryon density in the leading order has been already taken into account in Eq. (\ref{dN(NZ)/dp(a)}) by means of the exponent of the distribution of observable nucleons. All other possible dependences are ignored in Eq. (\ref{dN(NZ)/dp(a)}) because they require introduction of phenomenological functional dependencies rather than only phenomenological parameters. It is intuitively clear that the coalescence coefficient should decrease with increasing local temperature. Could happen that only such a temperature dependence can improve the reproduction of the transverse-momentum spectra because the contribution of the hot regions will be suppressed. A decrease of the coalescence coefficient with increasing $u_\mu p^\mu$, i.e. the fragment energy in the local rest frame of the fluid, can additionally improve the reproduction. It is clear that introduction of any additional tuning parameter, and moreover, an additional functional form, improves agreement with the data. However, we refrain from doing this in view of ambiguity of such modifications.

\section{Summary}
\label{Summary}

Within the 3FD model complemented by the coalescence model at the freeze-out stage
we have studied light-fragment production in Pb+Pb collisions at SPS energies 
and compared the obtained results with recently published data by the NA49 Collaboration 
\cite{Anticic:2016ckv}.  The simulations were performed with different 
equations of state---a purely hadronic EoS \cite{gasEOS}  
and two versions of the EoS involving the   deconfinement
 transition \cite{Toneev06}, i.e. a first-order phase transition  
and a smooth crossover one.

It is found that scenarios with the deconfinement transition \cite{Toneev06} are preferable 
for reproduction rapidity distributions  of deuterons and $^3$He  
in the considered energy range, except for the case of deuterons at 20$A$ GeV where the hadronic 
scenario is slightly preferable. 
At the same time the 
transverse-mass spectra  
are not in that nice agreement with
experimental data from Ref. \cite{Anticic:2016ckv}.  Moreover, different scenarios (with and 
without the deconfinement transition) fail approximately to the same extent. 
This is in spite of good reproduction of   
the proton midrapidity transverse-mass spectra (within all considered scenarios) in the low-$m_T$ range \cite{Ivanov:2013yla} 
which is relevant to the present light-fragment data. 
This problem could be cured by introduction of local-temperature and/or momentum dependence 
of the coalescence parameter
that suppresses high-momentum contributions to the fragment spectra. This modification implies 
introduction of additional fitting parameters that by itself extends the possibility of the data reproduction.  
The role of the afterburner stage of the collision , i.e. the hadronic cascade after the freeze-out, 
is another open question. It also can change the final fragment spectra.

It would be instructive to compare the coalescence results with those of 
the thermodynamical approach to the light-fragment production also based on the 3FD 
simulations. 
This question can be answered within the framework of recently developed 3FD event generator 
complemented by the Ultra-relativistic Quantum Molecular Dynamics (UrQMD) for the afterburner 
stage---a 
Three-fluid Hydrodynamics-based Event Simulator Extended by UrQMD final State interactions (THESEUS)
\cite{Batyuk:2016qmb}---because the thermal 
fragment production has been already incorporated in it \cite{Bastian:2016xna}.

\begin{acknowledgments}
This work was carried out using computing resources of the federal collective usage center «Complex for simulation and data processing for mega-science facilities» at NRC "Kurchatov Institute", http://ckp.nrcki.ru/.
Y.B.I. was supported by the Russian Science
Foundation, Grant No. 17-12-01427.
A.A.S. was partially supported by  the Ministry of Education and Science of the Russian Federation within  
the Academic Excellence Project of 
the NRNU MEPhI under contract 
No. 02.A03.21.0005. 
\end{acknowledgments}

\end{document}